\documentclass[trackchanges, twocolumn]{aastex701}
\usepackage{subcaption}
\usepackage{amsmath}

\begin{document}

\title[Two Stellar Populations in UBC 63]{Discovery of an Unbound Flyby Companion of UBC 63: In the Immediate Aftermath of a Close Encounter}

\author[orcid=0009-0005-2615-4547,sname='Biswas']{Samrat Biswas}
\affiliation{Department of Physics, Gauhati University, Gopinath Bordoloi Nagar, Jalukbari, Guwahati 781014 ,Assam,India}
\email[show]{samrat@gauhati.ac.in}  

\author[orcid= 0000-0002-3448-8150,sname='Medhi']{Biman J. Medhi} 
\affiliation{Department of Physics, Gauhati University, Gopinath Bordoloi Nagar, Jalukbari, Guwahati 781014 ,Assam,India}
\email[show]{bimanjmedhi@gmail.com}  

\author[orcid=0000-0002-2851-2468,sname='Messina']{Sergio Messina}
\affiliation{INAF—Catania Astrophysical Observatory, via S. Sofia 78, 95123 Catania, Italy}
\email{sergio.messina@inaf.it} 

\author[orcid= 0000-0002-6510-0681,sname='Zhu']{Zhanpeng Zhu}
\affiliation{Shanghai Key Lab for Astrophysics, Shanghai Normal University, Shanghai 200345, China}
\affiliation{Institute of Astronomy and Information, Dali University, Dali 671003, China}
\email{245514937@qq.com} 

\author[orcid= 0000-0003-3713-2640,sname='Songmei']{Songmei Qin}
\affiliation{Astrophysics Division, Shanghai Astronomical Observatory, Chinese Academy of Sciences, 80 Nandan Road, Shanghai 200030,
PR China}
\affiliation{School of Astronomy and Space Science, University of Chinese Academy of Sciences, No. 19A, Yuquan Road, Beijing 100049,
PR China}
\affiliation{Institut de Ciències del Cosmos, Universitat de Barcelona (ICCUB), Martí i Franquès 1, 08028 Barcelona, Spain}
\email{qinsongmei@shao.ac.cn} 

\author[orcid= 0000-0002-8833-5547,sname='Deb']{Sukanta Deb} 
\affiliation{Department of Physics, Cotton University, Guwahati 781001, Assam, India}
\email{sukantodeb@gmail.com}

\author[orcid=0009-0009-0399-0381,sname='Sheikh']{A. H. sheikh}
\affiliation{Department of Physics, Gauhati University, Gopinath Bordoloi Nagar, Jalukbari, Guwahati 781014 ,Assam,India}
\email{asheikh@gauhati.ac.in} 

\author[orcid=0000-0003-0262-7264,sname='Das']{H.S Das}
\affiliation{Department of Physics, Assam University, Silchar 788011, India}
\email{hsdas13@gmail.com} 

\author[orcid= 0000-0002-6510-0681,sname='Tamura']{Motohide Tamura}
\affiliation{Department of Astronomy, Graduate School of Science, The University of Tokyo, 7-3-1 Hongo, Bunkyo-ku, Tokyo 113-0033, Japan}
\affiliation{National Astronomical Observatory of Japan, National Institutes of Natural Sciences, Osawa, Mitaka, Tokyo 181-8588, Japan}
\email{motohide.tamura@nao.ac.jp} 

\author[orcid=0000-0002-1369-0608,sname='Gopinathan']{Maheswar Gopinathan}
\affiliation{Indian Institute of Astrophysics, Block II, Koramangala, Bangalore 560034, India}
\email{maheswar.g@iiap.res.in}

\author[orcid=0000-0003-4973-4745,sname='Sagar']{Ram Sagar}
\affiliation{Indian Institute of Astrophysics, Block II, Koramangala, Bangalore 560034, India}
\email{ram_sagar0@yahoo.co.in}

\begin{abstract}
We re-investigate the open cluster UBC 63 using the Gaia DR3 data and show that, rather than being a single cluster as previously classified, it is a compelling candidate for a double cluster undergoing an unbound flyby interaction. A GMM decomposition performed in the 5D astrometric space reveals the two statistically distinct components, namely UBC 63A (98 members, Age = 21 $\pm$ 4 Myr) and UBC 63B (148 members, Age = 562 $\pm$ 43 Myr). A significant age difference of $\Delta \mathrm{Age} = 541 \pm 43$ Myr between the clusters, rules out coeval formation. Their 3D separation of $60 \pm 29$ pc at the birth-epoch of the younger cluster, indicates that the clusters might have originated from the same molecular cloud complex. At present, the system exhibits a 3D separation of $26 \pm 8$ pc, with a relative velocity of $3.60 \pm 1.80$ km s$^{-1}$. Orbital integrations and \textit{N}-body simulations of the pair suggest that the systems had a close encounter, reaching a separation of $7 \pm 2$ pc only $\sim$~6 Myr ago and predict a rapid divergence to a separation of $491 \pm 213$ pc within the next $\sim$100 Myr. The low escape velocity ($V_{\rm esc} = 0.51 \pm 0.12$ km s$^{-1}$) of the system compared to the relative 3D velocity indicates that they are gravitationally unbound. Their low tidal factors, elongated structures and populations extending beyond the Jacobi radii may reflect a strong transient tidal interaction between the clusters.
\end{abstract}


\section{Introduction}
Open clusters (OCs) are gravitationally bound stellar systems that form within giant molecular clouds (GMCs) \citep{Lada_2003}. Most of them evolve as isolated entities. However, a small fraction forms in association with other clusters, giving rise to binary clusters (BCs) and, in some cases, higher-order systems \citep{2025A&A...693A.317Q}. The origin of binary and multiple cluster systems can be broadly grouped into five scenarios: simultaneous formation, sequential formation, tidal capture, resonant trapping, and optical pairing \citep{2009A&A...500L..13D}. These systems are astrophysically crucial because, unlike isolated clusters, they preserve signatures of mutual gravitational interaction, providing us with a rare opportunity to probe dynamical processes such as mass exchange, tidal distortion, and eventual merging or disruption \citep{2021ARep...65..755C, 2021A&A...649A..54P, 2022A&A...666A..75S}.

Before the \textit{Gaia} era, limited astrometric precision made membership determination in OCs challenging, thereby complicating the identification of BCs. Despite these limitations, early studies made important contributions to this field. For example, \cite{1989SvA....33....6P} identified 5 possible cluster groups from a sample of 66 OCs with known spatial and kinematic properties. \cite{1995A&A...302...86S} adopted a separation threshold of 20 pc to identify 18 candidate pairs of BC and estimated that BCs constitute $\sim$8 \% of the OC population in the Galaxy. Later, \cite{2009A&A...500L..13D} found that BCs account for $\sim$12 \% of OCs in our Galaxy. 

In recent years, high-precision astrometric and multi-band photometric data from \textit{Gaia} \citep{2018A&A...616A...1G, 2021A&A...649A...1G, 2023A&A...674A..41G} have substantially improved the study of BCs. \cite{2017A&A...600A.106C} performed the first systematic search for binary clusters and cluster groups based on 6D phase-space information: right ascension $(\alpha)$, declination ($\delta$), parallax $(\varpi)$, proper motions ($\mu^*_\alpha, \mu_\delta$) and radial velocities (RV). They identified 19 open cluster groupings, including 14 pairs, 4 groups with 3-5 members, and 1 complex with 15 members. Some very recent discoveries include the tidally captured pair NGC 1605a and NGC 1605b, with ages of $\sim$ 2 Gyr and 600 Myr, respectively \citep{2021ApJ...923...21C}; the primordial binary cluster pairs: NGC 2323-a and NGC 2323-b, aged at 158 Myr \citep{2025A&A...693A.317Q} and ASCC 19 and ASCC 21, with an estimated age of only $\sim$8.9 Myr \citep{2025AJ....169...98H}; the gravitationally unbound flyby encounter pairs: ASCC 71 and ESO 064-65 and NGC 2129 and UBC 437 \citep{2025A&A...702A.259Z}. Several other noteworthy BC candidates spanning a wide range of ages and formation histories have been discovered and studied using combinations of spatial proximity, velocity consistency, age similarity, shared membership, signatures of tidal interaction and \textit{N}-body simulations in the recent years \citep[e.g.,][]{2019ApJS..245...32L, 2021ARep...65..755C, 2022A&A...666A..75S}.  


In this study, we present evidence that the open cluster UBC 63, previously classified as a single system \citep[R.A. (J2000) = $05^h18^m41.6^s$, Dec. (J2000) = $+37^{\circ}48'31''$; $l = 169^{\circ}.5051$, $b = 0^{\circ}.1540$;][]{2023A&A...674A..37G}, is more likely a cluster pair undergoing an unbound flyby. UBC 63 has not been the subject of a dedicated, in-depth investigation in the existing literature.  Nevertheless, some previous works provide constrained estimates of its fundamental parameters, including $(\varpi)$: 0.626 to 0.692 mas, $\mu^*_\alpha$: 1.059 to 1.134 mas yr$^{-1}$, $\mu_\delta$: -3.516 to -3.563 mas yr$^{-1}$, RV $\sim$ -22.15 km s$^{-1}$, age $\sim$ 21.3 to 60.3 Myr, metallicity (z) $\sim$ 0.010 to 0.011 , distance $\sim$ 1470 to 1546 pc, extinction ($A_V$) $\sim$ 0.45 to 0.838 mag, Mass (M) $\sim$ 291 $M_\odot$ etc. \citep[e.g.,][]{2019A&A...627A..35C, 2020A&A...633A..99C, 2021MNRAS.504..356D, 2022A&A...668A...4F, 2023ApJS..268...46L, 2025A&A...693A.305A}. 

We perform a robust re-investigation of the system using the astrometric and photometric measurements from \textit{Gaia} Data Release 3 \citep[DR3;][]{2023A&A...674A..41G}. The cluster members for UBC 63 were retrieved using the membership analysis technique provided by \cite{deb_2022}. This sample was then decomposed using Gaussian mixture models (GMMs), into two statistically distinct stellar populations (UBC 63A and UBC 63B) with markedly different ages. We probe the dynamical evolution of the pair by means of orbital integrations and direct \textit{N}-body simulations. The results strongly favour the unbound flyby scenario with the closest approach occurring merely $\sim$6 Myr ago. The system also shows signatures of strong yet transient tidal interactions. We thus capture the system in a fleeting and rarely observed phase: an unbound cluster pair in the immediate aftermath of a close flyby passage. 



The remaining paper is organized as follows:  Section \ref{sec:discovery} describes the discovery of the cluster pair. The characterization of the two clusters is presented in sec. \ref{sec:characterization}. The past and future dynamical evolution are discussed in Sec. \ref{sec:orbit} and \ref{sec:nbody}. Sec. \ref{sec:summ} presents the summary of the work


\section{Discovery of the cluster pair}
\label{sec:discovery}

\begin{figure*}
\begin{subfigure}{0.66\columnwidth}
        \includegraphics[width=\columnwidth]{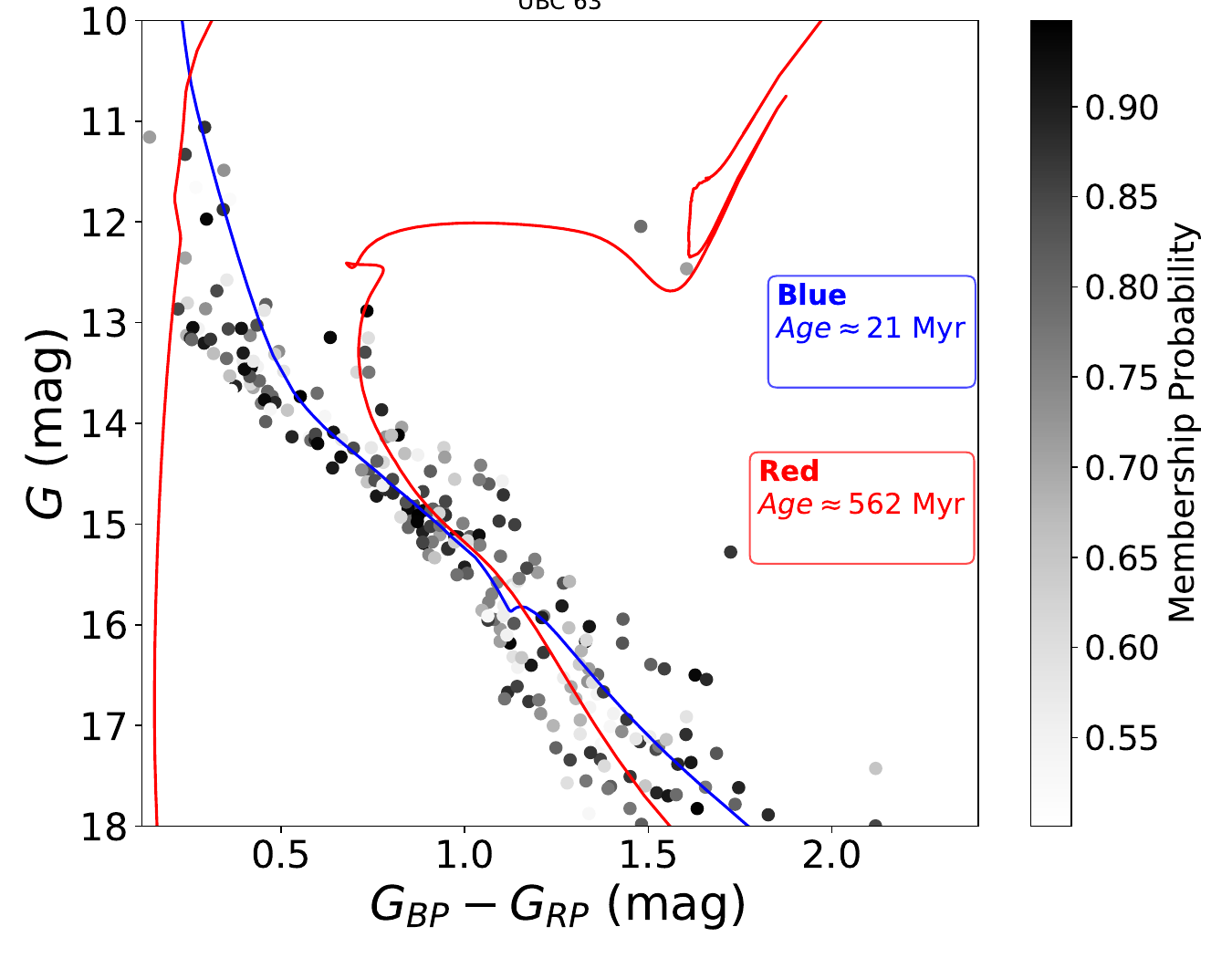}
        \caption{}
        \label{Fig:UBC63_iso}
\end{subfigure}
\begin{subfigure}{0.66\columnwidth}
        \includegraphics[width=\columnwidth]{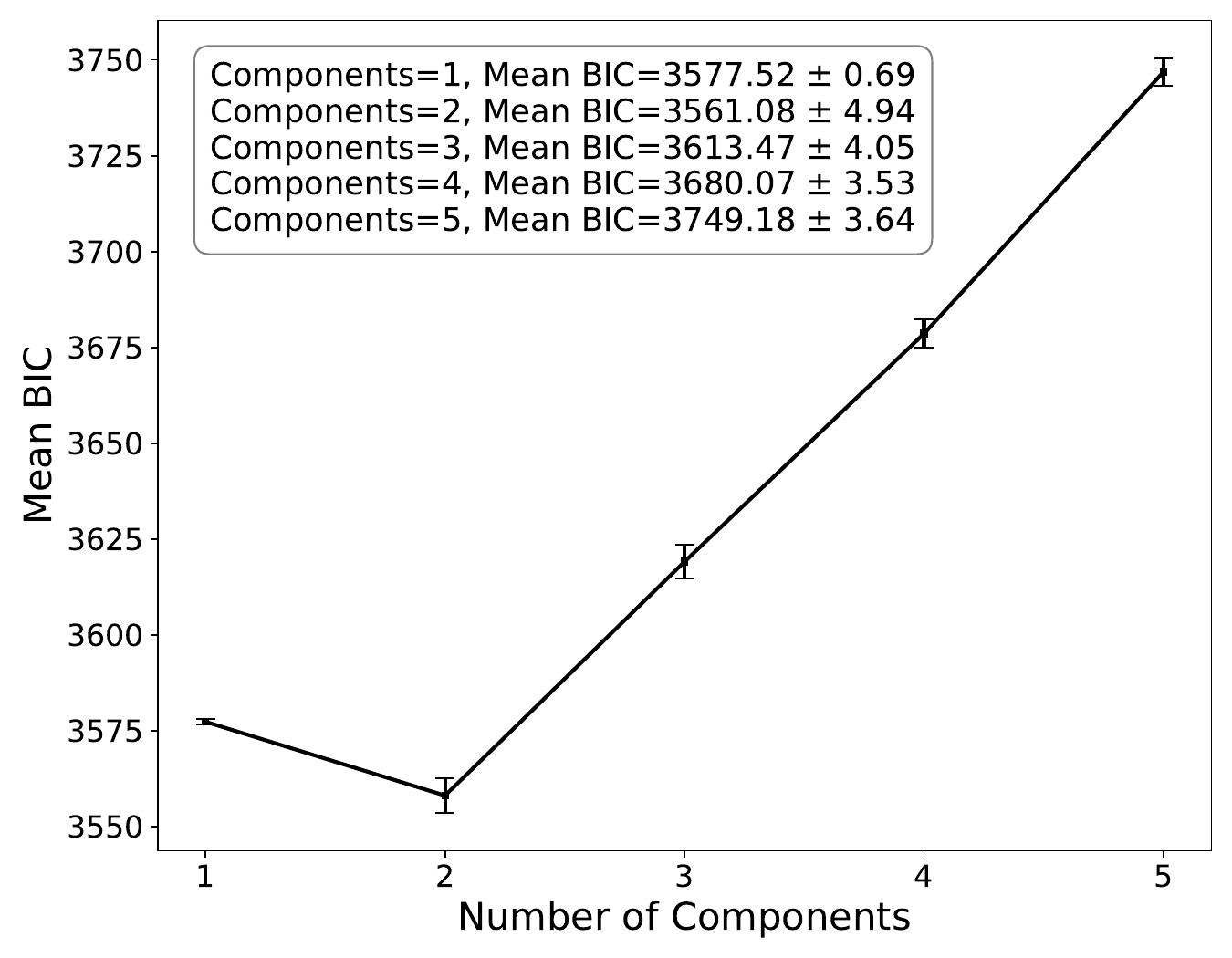}
        \caption{}
        \label{Fig:BIC}
\end{subfigure}
\begin{subfigure}{0.66\columnwidth}
        \includegraphics[width=\columnwidth]{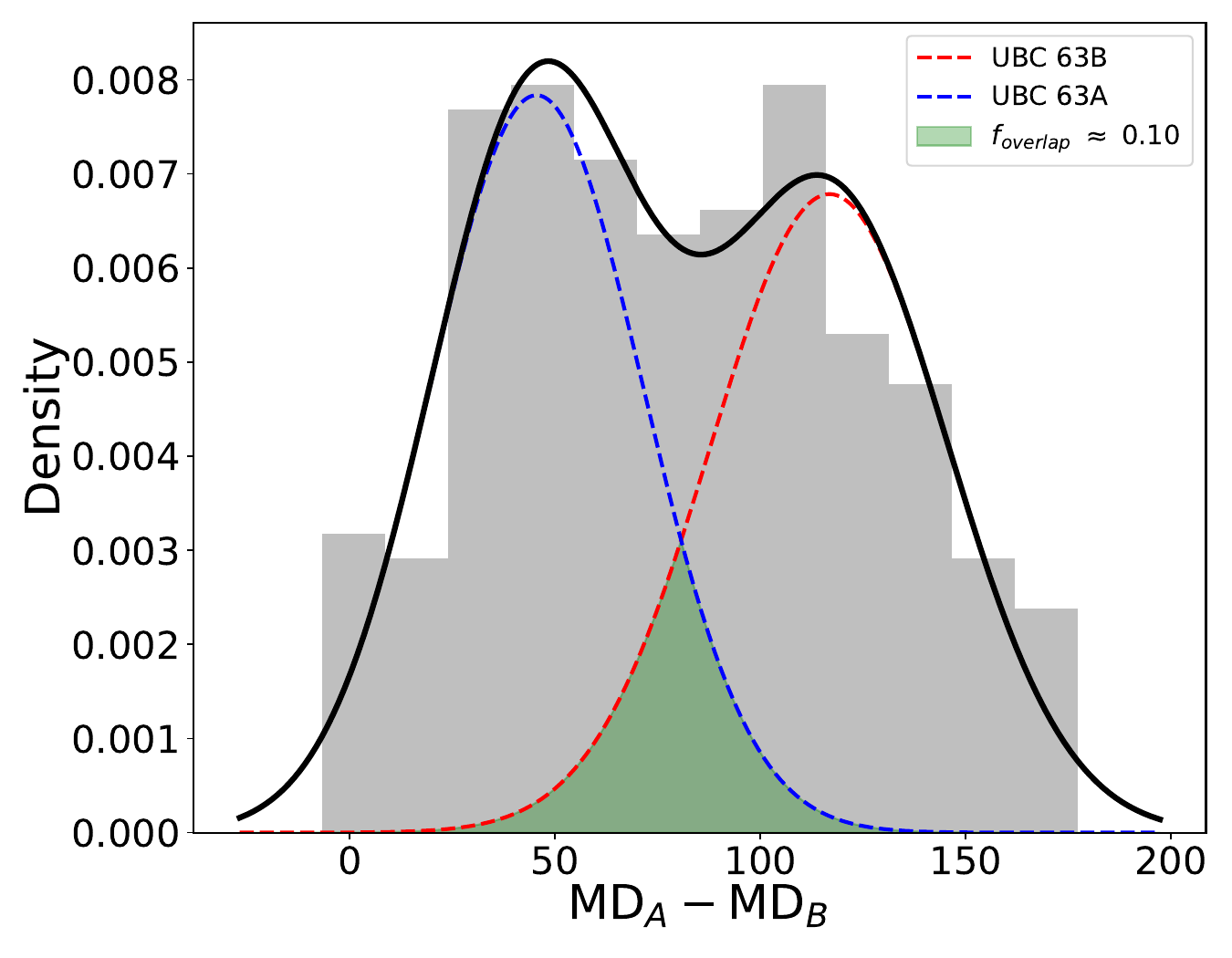}
        \caption{}
        \label{Fig:sep_MD_GMM}
\end{subfigure}\\
\begin{subfigure}{\columnwidth}
        \includegraphics[width=\columnwidth]{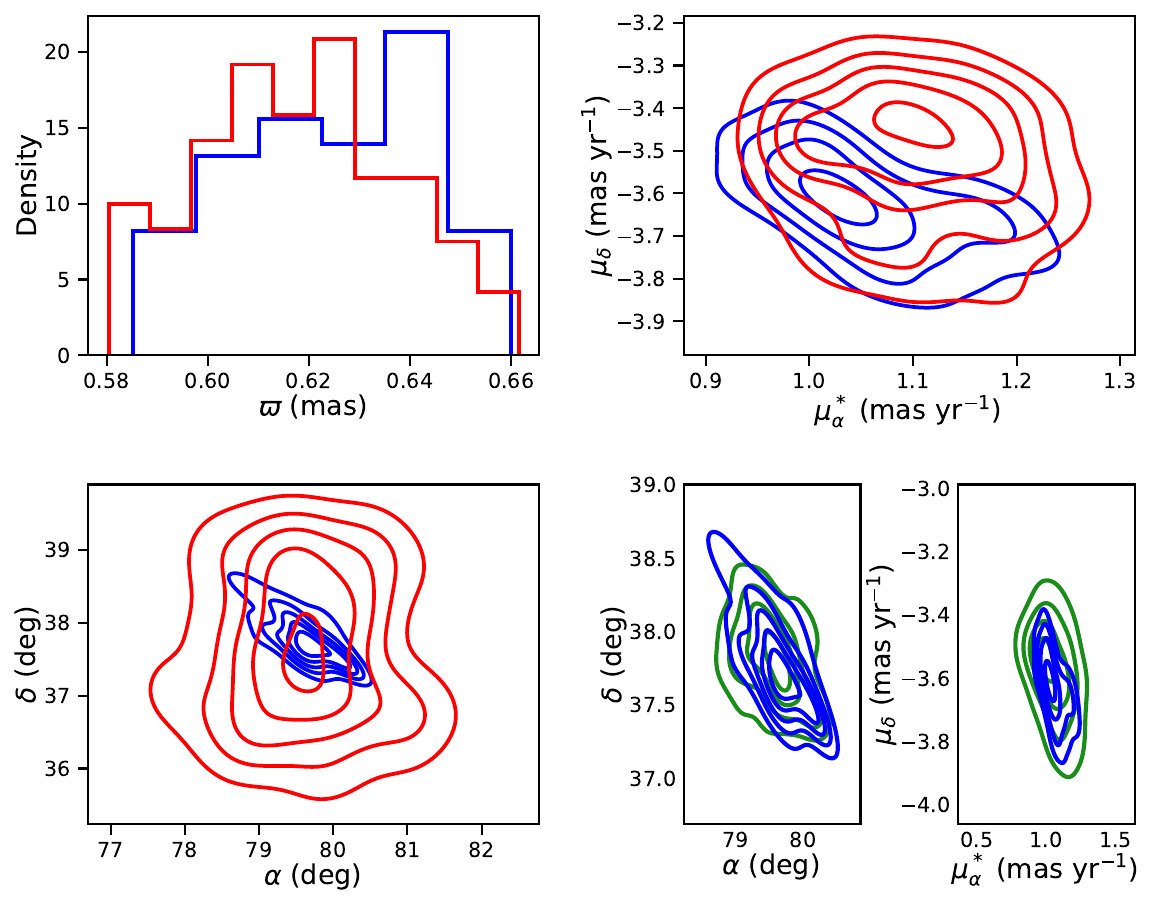}
        \caption{}
        \label{Fig:pm_dist}
\end{subfigure}
\begin{subfigure}{\columnwidth}
        \includegraphics[width=\columnwidth]{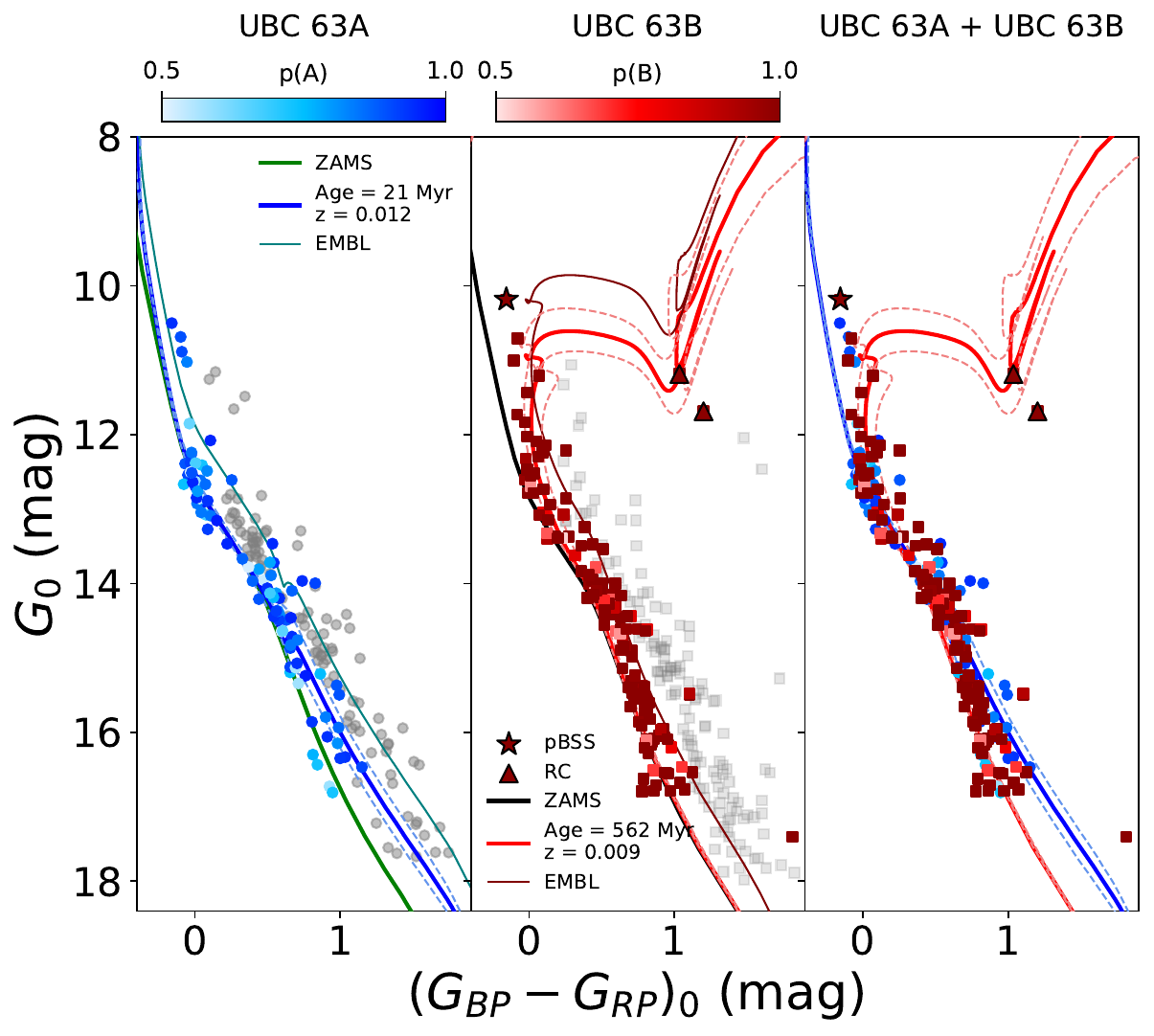}
        \caption{}
        \label{Fig:pop_sep_iso}
\end{subfigure}
\caption{(a) CMD of the 246 UBC 63 members with isochrones of 21 Myr (red) and 562 Myr (blue). (b) The BIC distribution for GMM models with 1–5 components obtained from 100 MC realizations. The mean BIC values along with their standard error for different number of components are also indicated. (c) Distribution of the difference in squared MD from the mean astrometric parameters of UBC 63A and UBC 63B, fitted with a two-component GMM. Blue and red curves denote UBC 63A and UBC 63B, respectively; the black curve shows the combined model, and the green shading indicates the overlap fraction. (d) Distributions of UBC 63A (blue) and UBC 63B (red) in parallax (upper left), proper-motion (upper right), and  $\alpha-\delta$ space (lower left). The lower-right panel shows the spatial and kinematic distributions of UBC 63A (blue) and classical UBC 63 (green). (e) Isochrone fits to the CMDs of UBC 63A (left), UBC 63B (middle), and the combined population (right). Points are color-coded by membership probability in UBC 63A (blues) and UBC 63B (reds); gray points represent the CMD points without reddening correction. The dashed curves represent the uncertainty bounds of the respective isochrone fits. The zero-age main sequence (ZAMS) and the equal-mass binary locus (EMBL) are included in the left and middle panels, as indicated in the legend. The best-fit age and metallicity ($z$) are listed in the upper right and lower-left corners of the left and middle panels. The red star and triangle symbols in the middle and right panels mark a probable blue straggler star (pBSS) and two red clump (RC) stars, respectively. }
\end{figure*}

Assuming UBC 63 to be a single cluster, we performed its membership analysis using astrometric data from \textit{Gaia} DR3 \citep{2018A&A...616A...1G, 2021A&A...649A...1G, 2023A&A...674A..41G}. The data was extracted within a 2$^{\circ}$ radius centered at R.A. (J2000) = $05^h18^m36^s$ and Dec. (J2000) = $+37^{\circ}37'18''$. We selected sources with $G < 18$ mag, $\varpi$ $>$ 0, RUWE $<$ 1.4, astrometric excess noise $<$ 2, visibility periods $>$ 10 and reliable five-parameter solutions. 

The membership determination was performed using a kNN-assisted Gaussian Mixture Model (GMM) approach. Approximate cluster astrometric ranges were first estimated from the central 10 arcmin region and used to pre-select stars across the 2$^{\circ}$ field. The resulting 3D astrometric distribution was transformed into a 1D Mahalanobis Distance (MD) distribution \citep{Mahalanobis_1936,DEMAESSCHALCK_2000} and modeled with a two-component (cluster and field) GMM. Membership probabilities were derived via an Expectation-Maximization (EM) algorithm, providing 246 members with $P>0.5$. Further methodological details are available in \cite{deb_2022, Biswas_2024, 2025MNRAS.541.1557B}.


Fig. \ref{Fig:UBC63_iso} presents the color-magnitude diagram (CMD) of the 246 probable members, color-coded according to their membership probabilities. The CMD morphology strongly suggests that a single isochrone cannot adequately reproduce the observed stellar distribution, as two clearly separated sequences are visible. This CMD feature is strikingly similar to the case of the old binary open cluster NGC 1605a and NGC 1605b reported in \cite{2021ApJ...923...21C}. As a preliminary inspection, we visually fitted two representative isochrones corresponding to ages of $\sim$21 Myr and $\sim$562 Myr. Thus, the presence of distinct sequences prompted a detailed investigation into a possible second cluster toward UBC 63. We note, however, that the CMD shown here has not been corrected for differential reddening, which may contribute additional scatter to the observed sequences and also bias the parameter estimates. Therefore, in sec. \ref{sec:characterization}, we examine de-reddened CMDs and derive the corresponding cluster parameters using the corrected photometry.


To investigate further and to achieve the population separation, we performed a probabilistic GMM in the scaled 5D astrometric space ($\alpha, \delta, \mu^*_\alpha, \mu_\delta, \varpi$) of the member stars. The Bayesian Information Criterion (BIC), used as the primary model-selection metric, was evaluated for models containing 1 to 5 components. The representative BIC values and their uncertainties were inferred from 100 Monte Carlo (MC) realizations, generated by sampling the astrometric parameters within their observational uncertainties. The BIC distribution showed a clear minimum for a two-component solution (See Fig. \ref{Fig:BIC}).  Thus, it was inferred that the distribution comprises two statistically distinct groups. Hereafter, we designate the two populations as clusters UBC 63A and UBC 63B, respectively. Based on a membership probability threshold of $>0.5$, we assigned 98 stars to UBC 63A (with $p(A)>0.5$) and 148 stars to UBC 63B (with $p(B)>0.5$) \footnote{A parallel analysis using a stricter threshold ($p(A/B)>0.7$ and $p (A/B)>0.9$) provided consistent results. However, we report the more inclusive ($p (A/B)>0.5$) sample for completeness.}. Fig.~\ref{Fig:sep_MD_GMM} shows their separation using a MD based discriminant, defined as $D = \mathrm{MD}_A - \mathrm{MD}_B$. The parameters $\mathrm{MD}_A$ and $\mathrm{MD}_B$ are squared MDs of each star from the 5D mean parameters of UBC 63A and UBC 63B, respectively. The negative and positive values of $D$ represents stars preferentially associated with UBC 63A and UBC 63B, respectively. The two-component GMM (blue for UBC 63A and red for UBC 63B) shows a low overlap fraction $(f{\mathrm{overlap}}) = 0.096$, indicating moderate but significant separation between the populations. Fig.~\ref{Fig:pm_dist} presents the distribution of UBC 63A (blue) and UBC 63B (red) in parallax (upper left), proper motion (upper right) and $\alpha - \delta$ space (lower left). The apparent overlap shows the similarity in the bulk kinematics of the two populations. A comparable spatial and kinematic similarity was also seen for the binary cluster pairs: NGC 1605a \& NGC 1605b in \cite{2021ApJ...923...21C}; Collinder 135 \& UBC 7 and UBC 547 \& UBC 549 in \cite{2022A&A...666A..75S}. This kinematic similarity might be an indication that the clusters were born within the same environment, although not at the same time. However, it should be noted that clusters at similar Galactocentric radii in the same spiral arm can share similar kinematics even without a common parent molecular cloud.  


In the lower right panel of Fig. \ref{Fig:pm_dist}, we show that the members of UBC 63A (blue) show near complete spatial/kinematic overlap with the previously reported members of the cluster UBC 63 (green) taken from \cite{2023A&A...675A..68V}. This shows that UBC 63A is same as the classical single cluster population of UBC 63. On the other hand, UBC 63B, although partially overlapping, exhibits systematic offsets in proper motion and parallax relative to UBC 63A. Furthermore, UBC 63B also forms a coherent kinematic structure (shown in Fig. \ref{Fig:pm_dist} in red ) with a well-defined CMD sequence (Fig. \ref{Fig:pop_sep_iso}, in red), strongly disfavoring a field-star origin. Its distinct age as compared to UBC 63A (see sec. \ref{sec:characterization}) further rules out an extended halo interpretation \citep{2022AJ....164...54Z}. 



\section{Characterization of the two populations}
\label{sec:characterization}

\begin{table*}
\centering
\caption{Fundamental parameters of UBC 63A and UBC 63B.}
\resizebox{1.15\textwidth}{!}{
\begin{tabular}{lcccccccccc}
\hline
\hline
Population & $\alpha$ & $\delta$ & $\varpi$ & $\mu^*_{\alpha}$ & $\mu_{\delta}$ & Age & $A_V$  & $d_{\rm bayes}$ & z & RV \\
 & (deg) & (deg) & (mas) & (mas\,yr$^{-1}$) & (mas\,yr$^{-1}$) & (Myr) & (mag)  & (pc) &  & (km\,s$^{-1}$) \\
\hline
UBC 63A & $79.599 \pm 0.042$ & $37.800 \pm 0.033$ & $0.624 \pm 0.002$ & $1.061 \pm 0.008$ & $-3.632 \pm 0.011$ & $21 \pm 4$ & $1.148 \pm 0.007$ &  $1502 \pm 5$ & $0.012 \pm 0.003$ & $-17.005 \pm 0.99$ \\
UBC 63B & $79.649 \pm 0.076$ & $37.666 \pm 0.080$ & $0.618 \pm 0.002$ & $1.099 \pm 0.006$ & $-3.521 \pm 0.012$ & $562 \pm 43$ & $1.025 \pm 0.011$ & $1527 \pm 5$ &  $0.009 \pm 0.002$ & $-13.54 \pm 1.38$\\
\hline
\end{tabular}
}
\setlength{\leftskip}{-70pt}\\
\setlength{\leftskip}{0pt}
\small{Notes: $\alpha$: R.A.; $\delta$: Dec.; $\varpi$: Parallax; $\mu^*_{\alpha}$, $\mu_{\delta}$: Proper motion in R.A. and Dec.; $A_V$: Visual extinction; $d_{bayes}$: Distance; $z$: Metallicity; $\text{RV}$: Radial velocity}  
\label{Tab:cluster_param}
\end{table*}


The mean cluster positions and proper motions of UBC 63A and UBC 63B were derived by modelling their astrometric distributions with independent Gaussian profiles. The Radial velocities (RVs) were estimated by cross-matching their member stars with the Survey of Surveys (SoS) catalog \citep{Tsantaki2022A&A...659A..95T}. This provided measurements for 9 and 18 stars in UBC 63A and UBC 63B, respectively. Mean RVs for each system were derived using inverse-variance weighting with iterative $3\sigma$ clipping \citep{2025A&A...693A.317Q} and the uncertainty was estimated following \cite{Soubiran2013A&A...552A..64S}. The difference in RVs of the two systems is calculated as $\Delta RV = 3.46 \pm 1.70$ km s$^{-1}$, which again indicates that the two systems are kinematically distinct. The distances ($d_{bayes}$) to UBC 63A and UBC 63B, were derived using the photo-geometric estimates from \cite{2021AJ....161..147B}. The mean cluster extinctions were inferred from their member star distributions in the $A_V$ space using values from \cite{2022A&A...658A..91A}. To constrain the age and metallicity ($z$), we constructed de-reddened Color Magnitude Diagrams (CMDs) for each cluster (Fig.~\ref{Fig:pop_sep_iso}). The de-reddening was performed on a star-by-star basis using the extinction values from \cite{2022A&A...658A..91A}. Although this approach has some dependency on the adopted extinction estimation, it is still  necessary to mitigate the differential reddening, which otherwise broadens the CMD and may bias the inferred cluster parameters \citep{2024MNRAS.532.2860R}. The isochrone fitting was carried out using the \texttt{isochrones}\footnote{\url{https://isochrones.readthedocs.io/en/latest/}} package with MIST v1.2 models \citep{2016ApJS..222....8D, 2016ApJ...823..102C, 2026ApJS..283...41B, 2026ApJS..283...64D}. All the estimated cluster parameters are listed in table~\ref{Tab:cluster_param}.


The age and metallicity difference between UBC 63A and UBC 63B is computed as $\Delta \mathrm{Age} = 541 \pm 43$ Myr and $\Delta \mathrm{z} = 0.003 \pm 0.004$. The significant age difference reinforces the interpretation that the two populations are distinct and also effectively rules out the scenario of simultaneous formation. The metallicity difference is not significant enough (considering the uncertainties) to confirm chemical distinction between the populations.





The star symbol shown in the middle and right panels of Fig.~\ref{Fig:pop_sep_iso} is a member of UBC 63B located above the main-sequence turn-off. It possesses a GALEX near-ultraviolet (NUV) counterpart \citep{2009Ap&SS.320...11B} and exhibits a phase-folded double-wave pattern in its TESS light curve, characteristic of an ellipsoidal binary. The projected rotational velocity ($v \sin i \sim 79.45 \pm 8.04$ km s$^{-1}$) supports a rapidly rotating star \citep{2025A&A...698A.300S}. These properties suggest that this star can be considered as a probable BSS (pBSS) in a close binary system, which are typically found in clusters older than about 300 Myr \citep{2021MNRAS.507.1699J, 2025ApJ...989...16S}.


We also note that the two stars located in the red giant branch (RGB) region of the CMD of UBC 63B, marked using triangles in Fig. \ref{Fig:pop_sep_iso}, have been previously reported in the literature as red-clump (RC) stars \citep{2020MNRAS.495.3087L,2025A&A...701A.270K}. These two RC stars provide an important constraint on the older age of UBC 63B. Hence, the effect of astrometric uncertainties on their population assignment must be considered. For this, we repeated the population separation analysis (as described in the previous section) for 100 MC realizations, sampling the astrometric parameters within their observational uncertainties. Across the 100 MC runs, the two RC stars had mean probabilities of 0.62 $\pm$ 0.03 and 0.59 $\pm$ 0.04 of belonging to UBC 63B. Thus, according to our adopted membership criterion of $p$ $>$ 0.5, both RC stars remain confidently assigned to UBC 63B despite the astrometric uncertainties.

Thus, it is safe to say that the presence of these core-helium-burning stars independently supports the relatively advanced age of UBC 63B.

\section{Galactic trajectories}
\label{sec:orbit}

\begin{table*}
\centering
\caption{Properties of UBC 63A and UBC 63B used to generate mock clusters for N-body simulations}
\resizebox{1.15\textwidth}{!}{
\begin{tabular}{lcccccccccccc}
\hline
\hline
Population & $M$ & $(X, Y, Z)$ & $(U, V, W)$ & $W_0$ & $f_b$ & $r_{\rm hm}$ & $Q$ & $M_{\rm min}$ & $\alpha_l$ & $M_t$ & $\alpha_h$ & $M_{\rm max}$ \\
 & ($M_\odot$) & (kpc) & (km\,s$^{-1}$) &  &  & (pc) &  & ($M_\odot$) &  & ($M_\odot$) &  & ($M_\odot$) \\
\hline
UBC 63A & $294 \pm 58$ & $(-9.477 \pm 0.005,\ 0.274 \pm 0.001,\ 0.020 \pm 0.001)$ & $(22.03 \pm 0.97,\ 206.83 \pm 0.21,\ -1.77 \pm 0.07)$ & 1.018 & 0.29 & $9.44 \pm 1.73$ & 0.5 & 0.15 & $2.17 \pm 0.62$ & 1.40 & $-2.18 \pm 0.48$ & 5.20 \\
UBC 63B & $435 \pm 87$ & $(-9.502 \pm 0.005,\ 0.275 \pm 0.002,\ 0.019 \pm 0.002)$ & $(18.71 \pm 1.85,\ 207.54 \pm 0.36,\ -1.19 \pm 0.07)$ & 1.274 & 0.82 & $23.57 \pm 2.48$ & 0.5 & 0.11 & $2.81 \pm 0.45$ & 0.89 & $-1.96 \pm 0.36$ & 2.48 \\
\hline
\end{tabular}
}
\setlength{\leftskip}{-70pt}\\
\setlength{\leftskip}{0pt}
\small{Notes: $M$: Cluster mass; $X,Y,Z$: Galactocentric positions; $U,V,W$: Galactocentric velocity; $W_0$: King concentration parameter; $f_b$: Binary fraction;$r_{\rm hm}$: Half-mass radius; $Q$: Virial ratio; $M_{\rm min/ \rm max}$: Minimum/maximum stellar mass; $\alpha_{l/h}$: Low/high mass slope of mass function; $M_t$: Transition mass} 
\label{Tab:mcluster_params}
\end{table*}

We compute the Galactocentric positions and velocities of UBC 63A and UBC 63B using \texttt{astropy} \citep{2013A&A...558A..33A, 2018AJ....156..123A}. For the transformation, we adopt the Solar position and velocity as $(X,Y,Z)_\odot = (8,0,0.015)$ kpc and $(U,V,W)_\odot = (10,235,7)$ km s$^{-1}$ \citep{Bovy2015ApJS..216...29B}. The resulting positions and velocities are listed in Table \ref{Tab:mcluster_params}. The present 3D spatial separation between the two clusters is estimated as $\Delta D_{3D} = 26 \pm 8 $ pc and the total space velocities are estimated as $V_{\mathrm{3D},A} = 208.01 \pm 0.14$ km s$^{-1}$ and $V_{\mathrm{3D},B} = 208.40 \pm 0.21$ km s$^{-1}$. The corresponding relative 3D velocity is  $\Delta V_{\mathrm{3D}} = 3.60 \pm 1.80$ km s$^{-1}$. The derived $\Delta D_{3D}$ and $\Delta V_{\mathrm{3D}}$ are comparable to the case of the binary cluster pair ASCC 71 and ESO 064-05 in \cite{2025A&A...702A.259Z}.


To investigate the past and future trajectories of UBC 63A and UBC 63B, we integrated the orbits of the two systems both forward and backward in time using the Python package \texttt{galpy}\footnote{\url{https://docs.galpy.org/en/v1.11.2/}}, with the axisymmetric Galactic potential module MW{\textsc{Potential}}2014 \citep{Bovy2015ApJS..216...29B}. The orbits of both clusters were integrated forward in time for 100 Myr from the present epoch. The backward integrations were extended to their respective formation epochs, corresponding to -21 Myr for UBC 63A and -562 Myr for UBC 63B. In addition, we also used MC simulations to estimate the uncertainty in the trajectories. The past and future trajectories of UBC 63A (blue) and UBC 63B (red) in the $Z$-$R_{\mathrm{GC}}$ plane are shown in Fig.~\ref{Fig:orbit_RZ}. The dashed gray curves correspond to the 16th and 84th percentile bounds obtained from 500 MC simulations. It was found that both clusters follow low-eccentricity orbits with similar angular momenta, UBC 63A: $e = 0.084 \pm 0.025$, $L_x = -6.0 \pm 0.2$, $L_y = 13.8 \pm 0.7$, $L_z = 1940.3 \pm 1.5$ kpc km s$^{-1}$; UBC 63B: $e = 0.073 \pm 0.005$, $L_x = -5.6 \pm 0.4$, $L_y = 8.4 \pm 0.7$, $L_z = 1951.4 \pm 2.8$ kpc km s$^{-1}$. Such similar orbital eccentricity and angular momenta was also observed for the pair ASCC 71 and ESO 064-05 in \cite{2025A&A...702A.259Z}.



Fig. ~\ref{Fig:time_sep} presents the temporal evolution of the inter-cluster separation from –21 Myr to +100 Myr. The 3D separation between the two clusters at the formation epoch of UBC 63A (–21 Myr) is found to be $\Delta D^{\mathrm{birth}}_{3D} = 60 \pm 29$ pc. Within uncertainties, the inferred separation is comparable to the characteristic size of giant molecular clouds ($\sim$100 pc; \citealt{2025A&A...702A.259Z}), hinting at a possible common environment origin and potentially explaining the similar bulk kinematics of the two populations. The closest approach at $\sim$ - 6 Myr, reaching a minimum separation of $\Delta D^{\mathrm{min}}_{3D} = 7 \pm 2$ pc, likely reflects a transient phase alignment along their orbits. After the closest approach, the clusters are now diverging. Forward integration shows a steady increase in separation, reaching $\Delta D^{\mathrm{future}}_{3D} = 491 \pm 213$ pc at +100 Myr. This is consistent with a flyby-like encounter. The lack of recurrent close passages disfavors a bound configuration. 
We estimated the mutual escape velocity ($V_{\rm esc}$) of the pair using the cluster masses from Table~\ref{Tab:mcluster_params} and their present-day 3D separation, obtaining \(V_{\rm esc}=0.51\pm0.12\) km s\(^{-1}\). Accounting for the uncertainties, the relative 3D velocity (\(\Delta V_{\mathrm{3D}}\); Sec.~\ref{sec:characterization}) is found to substantially exceed \(V_{\rm esc}\). About 10,000 MC samplings of the masses, separation and relative velocity parameters, considering the uncertainties, provides a $\sim$ 98\% probability that the pair is gravitationally unbound.


Nevertheless, such unbound close encounters can still induce transient but non-negligible tidal effects \citep{2025A&A...702A.259Z}. To quantify this, we estimate their tidal factors (TFs) in the present day, using the expression in \citep{2025A&A...693A.218P}. The resulting values for UBC 63A and UBC 63B are 4.33 and 2.01 pc$^2/M_\odot$, respectively. These low values indicate that the clusters still experience strong tidal interactions \citep{2025A&A...702A..48L}. 



Fig.~\ref{Fig:tidal_tail} shows the spatial distributions of UBC 63A (blue) and UBC 63B (red) in a tangentially projected Cartesian plane \citep{2025A&A...704A.167S}. The arrows attached to the scatter points denote their proper-motion vector corrected for projection effects \citep{2009A&A...497..209V, 2025A&A...704A.167S}. The gray dashed line traces the cluster orbit and indicates the direction of motion. The black dashed circle marks the Jacobi limit \citep[][]{2024A&A...686A..42H} of each cluster. To highlight the overall morphology and degree of elongation, best-fitting ellipses are overlaid on UBC 63A and UBC 63B, shown in orange and green, respectively.



Fig.~\ref{Fig:tidal_tail} shows that both UBC 63A and UBC 63B possess a number of candidate members located beyond their respective Jacobi radii, many of which show projected motions broadly consistent with the orbital trajectories. UBC 63B appears substantially more diffuse than UBC 63A. This is expected from its older age \citep{2025A&A...704A..50J} and consequently longer exposure to Galactic tidal perturbations, allowing escaping members to disperse over a larger volume. Furthermore, the mutual perturbation during the close flyby encounter between UBC 63A and UBC 63B may have contributed to the stripping of stars from the young UBC 63A, while simultaneously acting as a catalyst to further scatter the already loosely bound, extended outer population of UBC 63B. A detailed investigation of the extended population is required to quantify the relative roles of the Galactic tidal field and the close flyby encounter in shaping the morphology of both clusters.


\section{NBODY Simulation}
\label{sec:nbody}
\begin{figure*}
\begin{subfigure}{0.57\columnwidth}
        \includegraphics[width=\columnwidth]{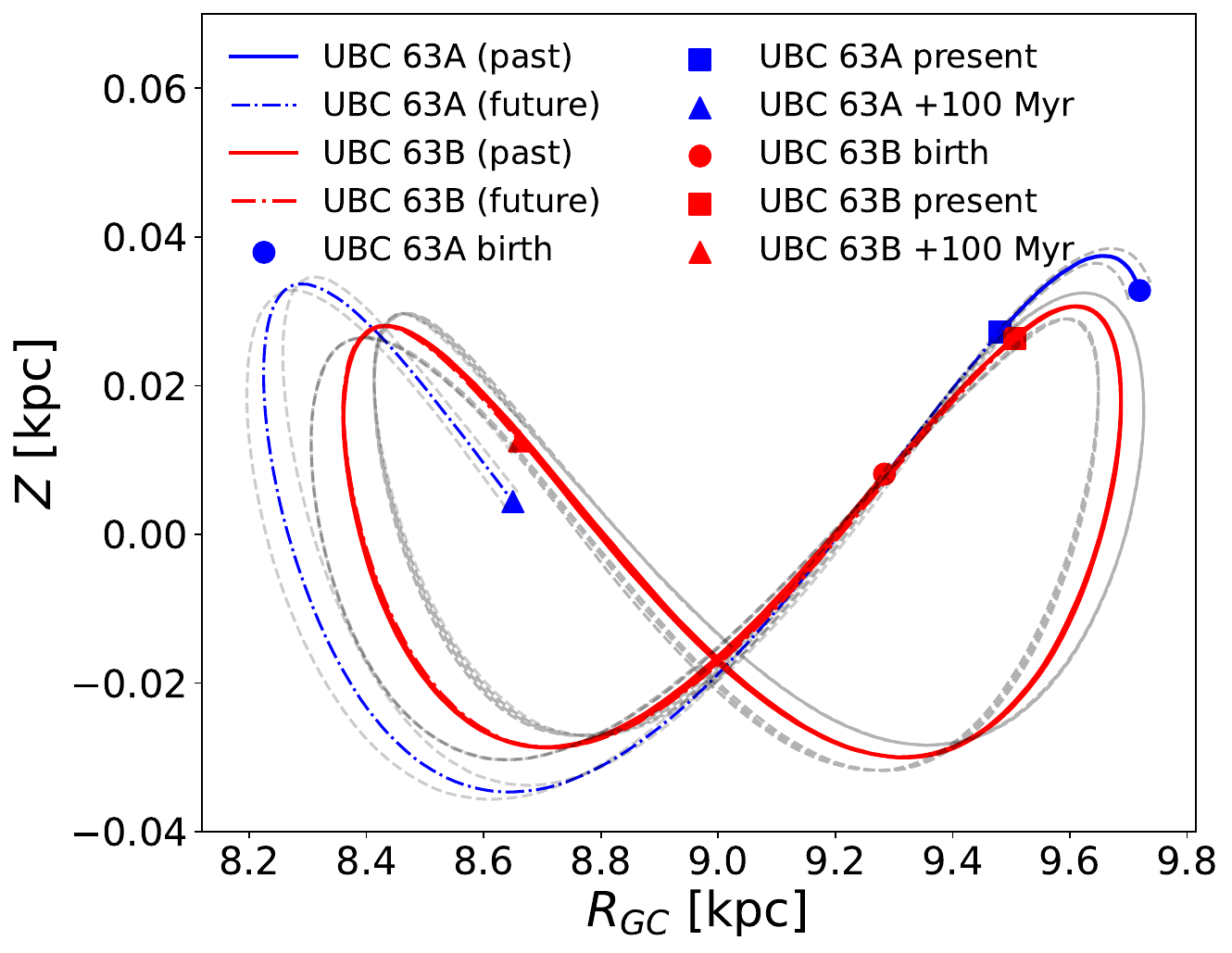}
        \caption{}
        \label{Fig:orbit_RZ}
\end{subfigure}
\begin{subfigure}{0.57\columnwidth}
        \includegraphics[width=\columnwidth]{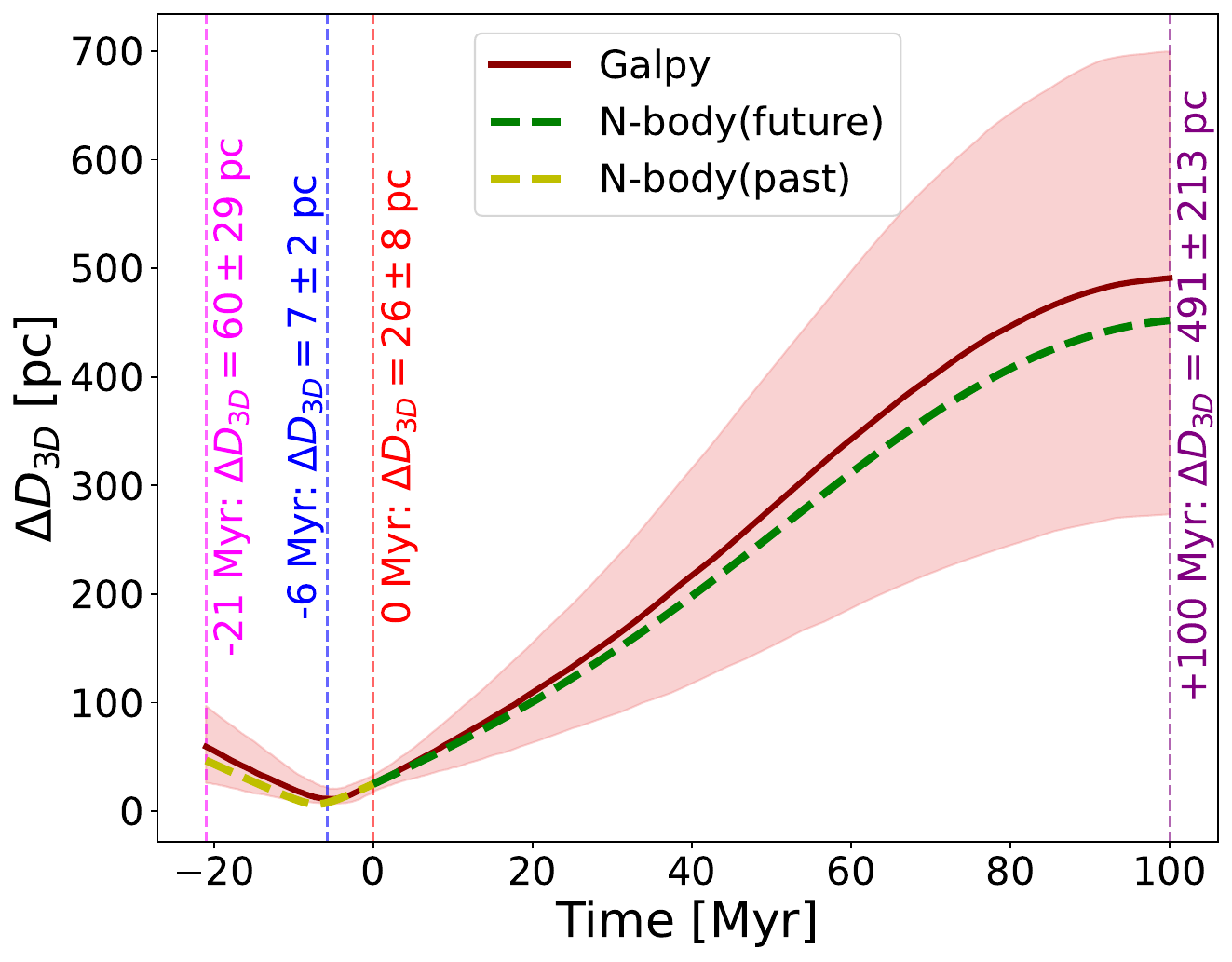}
        \caption{}
        \label{Fig:time_sep}
\end{subfigure}
\begin{subfigure}{0.89\columnwidth}\vspace{0.002cm}
\includegraphics[width=\columnwidth]{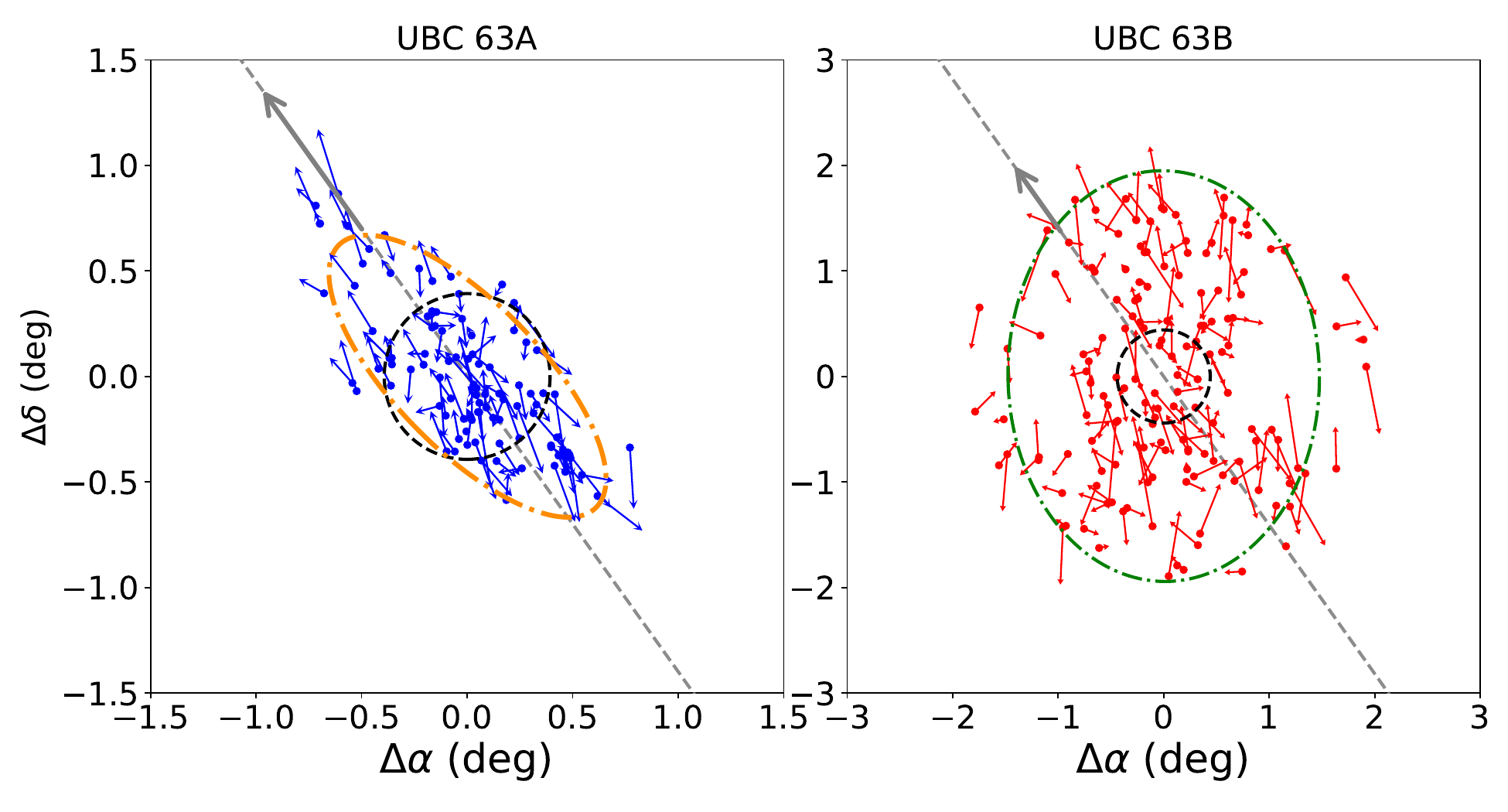}
\caption{}
\label{Fig:tidal_tail}
\end{subfigure}\\
\begin{subfigure}{0.66\columnwidth}
        \includegraphics[width=\columnwidth]{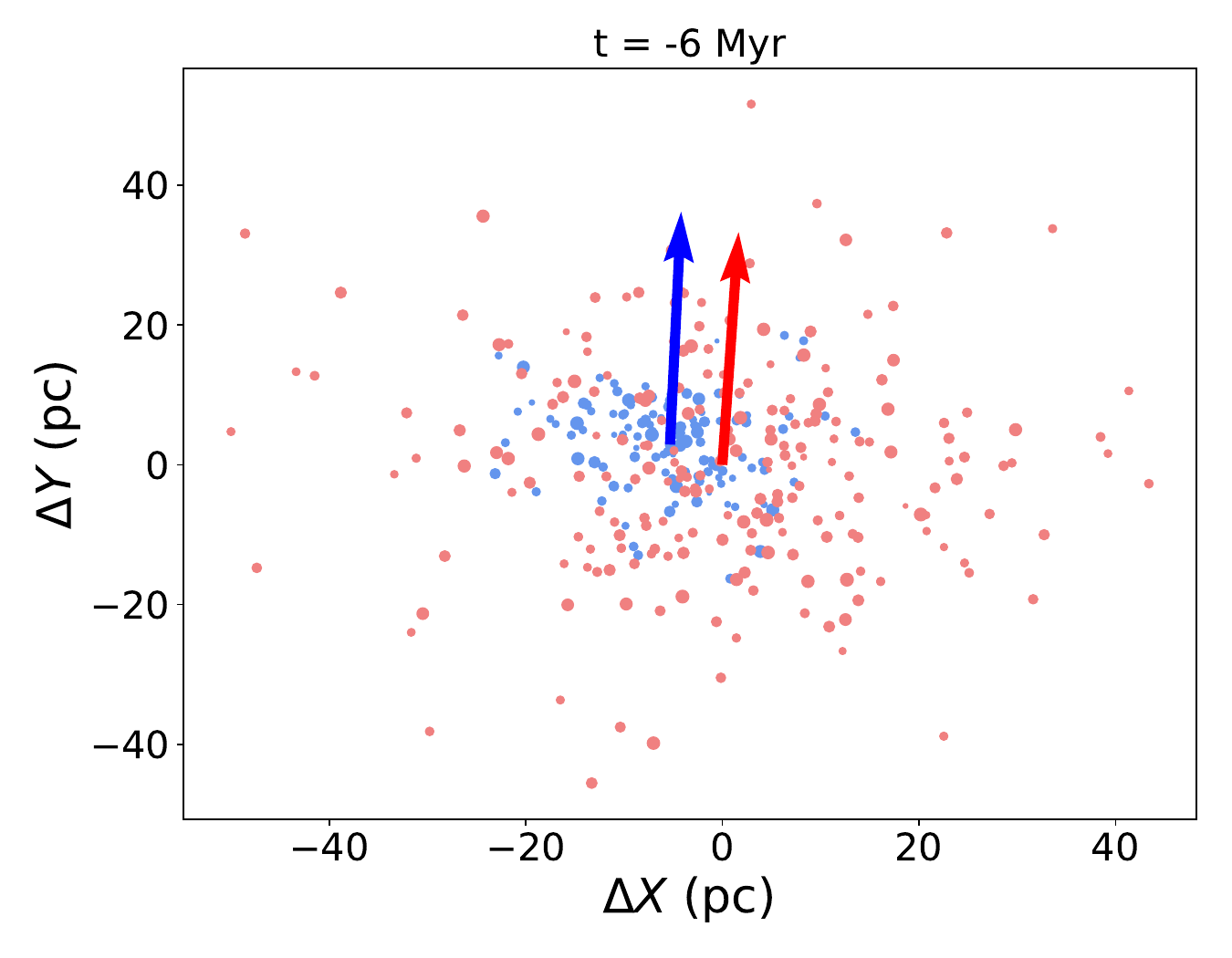}
        \caption{}
        \label{Fig:FW_UBC63_neg_6s}
\end{subfigure}
\begin{subfigure}{0.66\columnwidth}
        \includegraphics[width=\columnwidth]{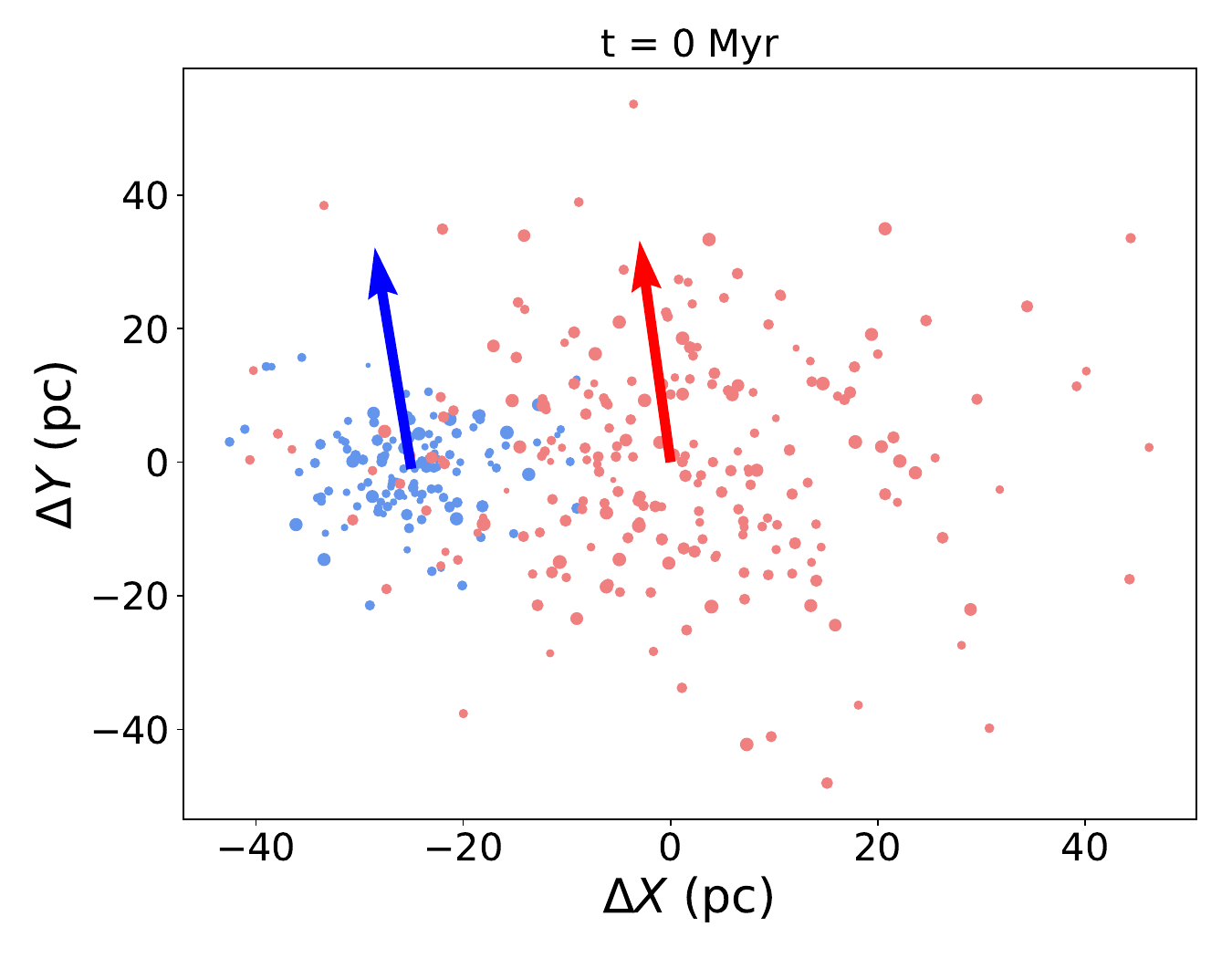}
        \caption{}
        \label{Fig:FW_UBC63_0}
\end{subfigure}
\begin{subfigure}{0.66\columnwidth}
        \includegraphics[width=\columnwidth]{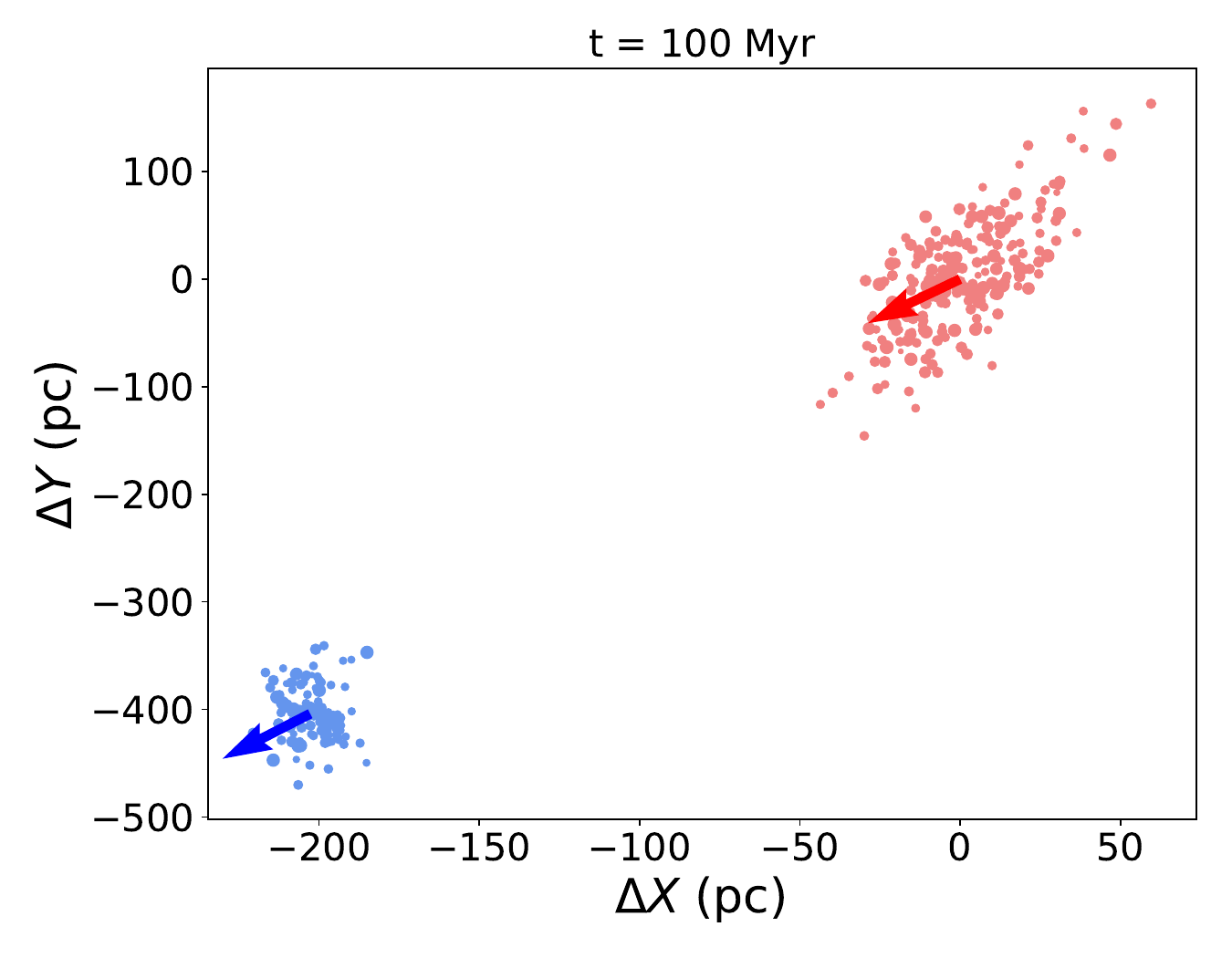}
        \caption{}
        \label{Fig:FW_UBC63_100}
\end{subfigure}
\caption{(a) The Galactic orbits of UBC 63A (blue) and UBC 63B (red) in the $Z$–$R_{\mathrm{GC}}$ plane. Solid and dash-dotted curves trace the past and future evolution, respectively. Circles, squares, and triangles mark the birth, present, and future positions (at +100 Myr), respectively. The dashed gray curves denote the 16th and 84th percentile uncertainty bounds from the MC sampling.  (b) Temporal evolution of the inter-cluster separation from –21 Myr to +100 Myr. The solid red curve shows the mean evolution derived from \texttt{galpy}. The corresponding shaded regions indicate the 16th and 84th percentile uncertainty bounds from MC sampling. The yellow and green dashed curves correspond to the past and future evolution scenarios from the \textit{N}-body simulation, respectively. The birth epoch of UBC 63A, closest approach, the present epoch and the +100 Myr epoch is marked using magenta, blue, red and purple dashed vertical lines, respectively, with the corresponding separations annotated. (c) The the spatial distributions of UBC 63A (blue) and UBC 63B (red) in a tangentially projected Cartesian plane. The arrows attached to the scatter points denote their first order corrected proper-motion vector. The gray dashed line traces the cluster orbit and indicates the direction of motion. The black dashed circle marks the Jacobi limit of each cluster. The best-fitting ellipses are overlaid on UBC 63A and UBC 63B, in orange and green, respectively. X-Y projection of the mock cluster of UBC 63A (blue) and UBC 63B (red) at 3 different epochs: (d) $-6$ Myr, (e) 0 Myr and (f) $+100$ Myr as obtained from the \textit{N}-body simulation. The blue and red arrows represent the space velocity projection of each cluster.}
\label{Fig:nbody}
\end{figure*}

The \texttt{galpy} orbital integration treats the clusters as point masses, an approximation that may sometimes oversimplify their dynamical behaviour \citep{2025A&A...702A.259Z}. Hence, to validate these result, we constructed mock clusters from the observational data using \texttt{McLuster}\footnote{\url{https://github.com/AlbrechtKamlah/mcluster}} \citep{2011ascl.soft07015K} and followed their evolution with the high performance \textit{N}-body code \texttt{PeTar}\footnote{\url{https://github.com/lwang-astro/PeTar}}\citep{2020MNRAS.497..536W}.

 As discussed in Section \ref{sec:characterization}, the Galactocentric positions $(X,Y,Z)$ and velocities $(U,V,W)$ of the two clusters were derived from the observed kinematics. Their cluster masses ($M$), binary fractions ($f_b$), minimum and maximum stellar masses ($M_{min/max}$), transition mass ($M_t$), and the low and high mass slopes ($\alpha_{l/h}$) of their segmented mass functions were obtained using the method of \cite{2023MNRAS.525.2315A}. The King concentration parameter ($W_0$) was derived from the $c-W_0$ correspondence provided in \cite{1966AJ.....71...64K}, where $c ~(= \rm tidal~radius/core~radius)$ was computed from \citep{1962AJ.....67..471K} fits to the radial density profiles (RDP) of the clusters. The half mass radius ($r_{hm}$) was computed using the expression from \citep{2006BaltA..15..547S}.


 We also estimated the dynamical relaxation time ($t_{relax}$) of the two clusters using the equation in \citep{binney2011galactic}. It was found to be $94 \pm 27$ Myr for UBC 63A and $415 \pm 77$ Myr for UBC 63B. While the age of UBC 63B exceeds its relaxation time, the younger UBC 63A is likely not yet dynamically relaxed. Nevertheless, for simplicity, we adopted a virial ratio of Q = 0.5 for both systems \citep{2025A&A...693A.317Q}. To assess the impact of this assumption, we additionally carried out PETAR simulations with sub-virial initial conditions (Q $<$ 0.5). We found that the overall results remain largely unchanged, indicating that the conclusions are not strongly sensitive to the assumed virial ratio.

All of these parameters were used as inputs to generate the mock clusters representing UBC 63A and UBC 63B. These are summarized in Table \ref{Tab:mcluster_params}.

The mock clusters were then evolved with \texttt{PeTar}. The simulations account for mutual stellar interactions \citep{2020MNRAS.497..536W}, single and binary stellar evolution \citep{2000MNRAS.315..543H,2002MNRAS.329..897H,2020A&A...639A..41B}, as well as the Galactic potential \citep{Bovy2015ApJS..216...29B}. Each system was first integrated forward for +100 Myr with snapshots recorded at 1 Myr intervals. 

To investigate the clusters' dynamical history up to the present day, we required backward integration, a capability which is not natively supported by \texttt{PeTar}. To resolve this, we first derived the phase space co-ordinates of both the clusters at $t = -21$ Myr, via reverse-time point-mass orbit integration in \texttt{galpy}. The clusters were then initialized at this epoch using these coordinates and assuming their structural properties to be same as the present day. The reconstructed system was then evolved forward to the present day. While this simplified approach is not expected to be perfectly accurate, it is sufficient for evaluating the dynamical consistency of the inferred interaction history \citep{2025A&A...693A.317Q}.

Fig.~\ref{Fig:FW_UBC63_neg_6s}, \ref{Fig:FW_UBC63_0} and \ref{Fig:FW_UBC63_100} presents the projected evolution of the cluster pair in the $X$–$Y$ plane at $-6$ Myr, the birth epoch of the younger cluster; $0$ Myr, the present day and $+100$ Myr, illustrating future evolution. The corresponding 3D spatial separations at these epochs are $\sim$ 7.4 pc, 25 pc and 452 pc respectively. These results agree well (within the error limits) with the \texttt{Galpy} orbital integration. The full temporal evolution of the separation, as obtained from the \textit{N}-body simulation is also presented in Fig.~\ref{Fig:time_sep}, as yellow (for past) and green (for future) dashed curves. 

Taken together, these results suggests that UBC 63A and UBC 63B are best interpreted as an unbound flyby pair, rather than a gravitationally bound binary or a tidally captured system.

\section{Summary}
\label{sec:summ}

In this study, we revisit the nature of UBC 63 and demonstrate that it is unlikely to be a single cluster, but instead a compelling candidate for a double cluster system (UBC 63A and UBC 63B) caught in the immediate aftermath of a close flyby. 

The clusters exhibit distinct evolutionary phases, with ages of $21 \pm 4$ Myr (UBC 63A) and $562 \pm 43$ Myr (UBC 63B) and metallicities of 0.012 $\pm 0.003$ (UBC 63A) and 0.009 $\pm 0.002$ (UBC 63B). The significant age disparity ($\Delta$ Age = $541 \pm 43$ Myr) effectively rules out coeval formation scenario. The older cluster, UBC 63B, also exhibits advanced stellar evolution signatures, hosting a probable blue straggler star (pBSS) as well as two red clump (RC) stars, consistent with its evolved status.

The dynamical orbit integration and direct \textit{N}-body simulations show that, at the formation epoch of UBC 63A, ($\sim$ –21 Myr), the two clusters were separated by $60\pm29$ pc. This suggests that the clusters may have originated from the same molecular cloud complex and also explains their similar kinematics. The closest approach occurred only $\sim$6 Myr ago, during which the separation reduced to $7 \pm 2$ pc. The system's low escape velocity ($V_{\rm esc} = 0.51 \pm 0.12$ km s$^{-1}$) compared to the relative 3D velocity $(\Delta V_{3D} = 3.60 \pm 1.80$ km s$^{-1}$) indicates that the pair is gravitationally unbound. At present, the system exhibits a 3D separation of $26 \pm 8$ pc. The future evolutionary scenarios of clusters show that they are now rapidly diverging and are expected to reach separations of $491\pm213$ pc within the next $\sim$100 Myr. 

The close flyby, occurring approximately 6 Myr ago, appears to have been dynamically significant and may have generated a transient but significant tidal interaction between the two clusters. This is consistent with their low tidal-factor values of 4.33 $M_\odot^{-1}$ pc$^{2}$ (UBC 63A) and 2.01 $M_\odot^{-1}$ pc$^{2}$ (UBC 63B). Both systems also display elongated morphologies and extended stellar populations reaching beyond their respective Jacobi radii, further suggesting ongoing tidal evolution.

Thus, the two clusters are best interpreted as an unbound tidally interacting flyby system. Thus, this system represents a rare, real-time laboratory for studying the tidal consequences of cluster-cluster interaction in the solar neighborhood. Understanding these brief but intense dynamical exchanges is crucial for refining our current models of open cluster survival, mass loss, and overall evolution in the solar neighborhood.

\section*{Acknowledgments}

The authors sincerely thank the anonymous referee for the constructive comments which significantly helped to improve the clarity and overall quality of the manuscript. This research has used the VizieR catalogue access tool, CDS, Strasbourg, France and  cross-match service provided by CDS, Strasbourg, observations made with the Galaxy Evolution explorer, obtained from the MAST data archive (available in the \dataset[doi: 10.17909/0gys-vy33]{\doi{10.17909/0gys-vy33}}) at the Space Telescope Science Institute, which is operated by the Association of Universities for Research in Astronomy, Inc., under NASA contract NAS 5–26555. SB and BJM acknowledge the IUCAA, Pune, for providing access to the Pegasus High Performance Computing (HPC) facility. SB highly acknowledges the Department of Science and Technology (DST), Govt. of India for providing DST INSPIRE fellowship vide grant no. IF220155. AHS also acknowledges the Department of Science and Technology (DST), Govt. of India, for providing the DST INSPIRE fellowship vide grant no. IF230175. The authors extend their sincere gratitude towards Annapurni Subramaniam, Gabriel I Perren, Pavel Kroupa, Vikrant Jadhav, Hektor Monterio and Ranit Behera for their valuable suggestions and help during this work. 

\bibliography{UBC63_refer}{}
\bibliographystyle{aasjournalv7}



\end{document}